\documentclass[aps,pre,twocolumn,groupedaddress,superscriptaddress]{revtex4}
\usepackage{amsmath}\usepackage{xcolor}\usepackage{epstopdf}
\usepackage{graphicx}
\usepackage{color,ulem}

\draft 
\begin{document}


\title{Optimizing optical tweezing with directional scattering in composite microspheres} 


\author{R. Ali}
\email[]{r.ali@if.ufrj.br}
\affiliation{%
Instituto de F\'isica, Universidade Federal do Rio de Janeiro, Caixa Postal 68528, Rio de Janeiro, RJ, 21941-972, Brasil
}
\author{F. A. Pinheiro}
\affiliation{%
Instituto de F\'isica, Universidade Federal do Rio de Janeiro, Caixa Postal 68528, Rio de Janeiro, RJ, 21941-972, Brasil
}
\author{ F. S. S. Rosa}
\affiliation{%
Instituto de F\'isica, Universidade Federal do Rio de Janeiro, Caixa Postal 68528, Rio de Janeiro, RJ, 21941-972, Brasil
}
\author{ R. S. Dutra}
\affiliation{LISComp-IFRJ, Instituto Federal de Educa\c{c}\~ao, Ci\^encia e Tecnologia, Rua Sebasti\~ao de Lacerda, Paracambi, RJ, 26600-000, Brasil}
\author{P. A. Maia Neto}
\affiliation{%
Instituto de F\'isica, Universidade Federal do Rio de Janeiro, Caixa Postal 68528, Rio de Janeiro, RJ, 21941-972, Brasil
}



\begin{abstract}
Trapping of microspheres with a single focused laser beam is usually limited to materials with relative refractive indexes slightly larger than one. 
We show that directional light scattering can be employed to optically trap high-index materials.
For this purpose, we propose a material platform to achieve  zero backward scattering (ZBS), also known as the first Kerker condition, 
 in a composite media containing spherical inclusions of 
silica embedded in a SiC microsphere. By tuning the volume filling fraction of inclusions and the microsphere radius, stable trapping can be achieved, provided that 
ZBS is combined with the condition for destructive interference between the fields  reflected at the external and internal interfaces of the 
microsphere when located at the focal point. 
We show that our proposal also holds even in the presence of 
a significant amount of spherical aberration, which is a common condition in most optical tweezers setups. 
In this case, achieving ZBS is essential for trapping high-index materials.

\end{abstract}

\maketitle

\section{Introduction}

Optical tweezers - laser traps for neutral microscopic particles - are a powerful optical tool with many applications in physics and biology~\cite{ashkin1986,ashkin2006}. In biology, optical tweezers 
allowed for pioneering quantitative measurements of basic interactions
in living cells~\cite{fazal2011}. In physics, optical tweezers have been employed to tackle important fundamental problems, such as the experimental implementation of the Szilard's
demon~\cite{toyabe2010}, the first experimental proof of Landauer's
principle~\cite{brut2012}  
and femtonewton force measurements that open the way for probing non-trivial geometry effects in the double-layer and Casimir interactions~\cite{diney}.  

 In a typical optical tweezer setup, a dielectric microsphere inside a water-filled sample chamber is illuminated  by a single laser beam  brought to a diffraction limited
focal spot.
Whereas reflection gives rise to radiation pressure that pushes the microsphere along the overall propagation direction, refraction 
provides  a restoring force pointing to the focal point if the sphere relative refractive index $m$ is larger than one. 
Trapping is achieved provided  $m$ is only slightly  larger than one, so that refraction dominates over reflection.
This condition imposes a severe limitation in many applications of practical interest. For instance, in Casimir force measurements with optical tweezers, 
the magnitude of the interaction usually increases with the sphere refractive index, and a larger experimental signal would be obtained if optical trapping of spheres with large refractive indexes was possible.

In order to circumvent this limitation we explore novel mechanisms to control the radiation scattering pattern, which is now possible due to progress in the field of metamaterials. Specifically, we introduce the concept of Zero Backward Scattering (ZBS), also known as first Kerker condition, in the field of optical  trapping. The first and the second Kerker conditions, the latter being related to Zero Forward Scattering (ZFS), were put forward in 1983, when it was theoretically shown that in a magnetic sphere coherent effects between electric and magnetic dipoles may lead to strongly asymmetric radiation patterns~\cite{ker}. For many years the observation of these effects for optical frequencies was considered to be impossible due to the fact that the relative magnetic permeability of natural media is unity in this frequency range. However, with the advent of metamaterials, which allowed for optical magnetism~\cite{monticone}, this limitation has been overcome and achieving directional scattering and Kerker conditions  have attracted considerable attention in recent years. Indeed, several experimental realizations to obtain directional scattering exist, such as single GaAs~\cite{Si}, silicon~\cite{fu2013,ge}, or dielectric nanoparticles~\cite{geffrin2012}, as well as gold nanoantennas~\cite{alaee2015}. In addition, different approaches toward directional and/or anomalous scattering patterns have been proposed, such as exploiting interferences among dipolar and quadrupolar resonances~\cite{li2015}, using gain~\cite{xie2015}, core-shell~\cite{liu2012} and highly refractive nanoparticles~\cite{kuznetsov2016}, or designing pyramidal nanostructures to excite magnetic resonances~\cite{rodriguez2014}. 

In the present paper we propose a new material platform to achieve ZBS based on non-magnetic composite media and apply it to facilitate and optimize optical tweezing. Using extended Maxwell-Garnett effective medium and Mie theories, we demonstrate that one can achieve nearly ZBS for some values of the the filling fraction of inclusions. For concreteness, we consider a microsphere of ${\rm SiC}$ with ${\rm SiO_2}$ inclusions at the typical operation wavelength of optical tweezers, 
$1064\,{\rm nm}.$ At such wavelength, an homogeneous microsphere of ${\rm SiC}$ cannot be optically trapped in water due to its large refractive index $n_{\rm SiC}=2.57$. We show that one can circumvent this limitation by adding inclusions that allow for a drastic reduction of backscattering radiation, hence optimizing optical trapping. We demonstrate that by fulfilling the first Kerker condition one can not only trap high refractive index spheres that cannot be trapped in general, but also enhance trapping stability. Besides, we prove that this result is even more important
if one considers  the spherical aberration caused  by refraction at the interface between the glass slide and the water-filled sample chamber,
which typically occurs in the vast majority of cases of practical interest. Altogether our results unveil the role of backscattering in optical tweezing and pave the way for the design of new experimental trapping devices.         

This paper is organized as follows. Section II is devoted to the description of the methodology, where Mie and extended Maxwell-Garnett theories are briefly presented. In Sec. III the conditions for optical tweezing are discussed and our main results are presented and analyzed. Finally, in Sec. IV we summarize our findings and conclusions.

\section{Methodology}

 \subsection{Mie Theory and Kerker conditions }

When an electromagnetic plane wave (vacuum wavelength $\lambda_0$) 
illuminates a spherical particle (radius $r$ and relative refractive index $m$), scattering and absorption processes can be described in terms of the spherical multipole decomposition using Mie theory~\cite{bohren}. The scattered field can be written as a function of the Mie coefficients $ a_{\ell}$ and $ b_{\ell}$, 
which correspond to the electric and magnetic multipoles, respectively. Here the index ${\ell}$ is used to denote the ${\ell}^{th}-$ order spherical harmonic channel.
The Mie coefficients are functions of the size parameter 
$x=kr,$ where $k$ is the wavenumber in the host medium \cite{bohren}:

\begin{equation}\label{aell}
 a_{\ell}= \dfrac{m\psi_{\ell}(mx) \psi'_{\ell}(x)-\mu\psi_{\ell}(x) \psi'_{\ell}(mx)}{m\psi_{\ell}(mx) \xi'_{\ell}(x)-\mu\xi_{\ell}(x) \psi'_{\ell}(mx)} 
 \end{equation}
 
\begin{equation}
\label{bell}
 b_{\ell}=  \dfrac{\mu \psi_{\ell}(mx) \psi'_{\ell}(x)-m  \psi_{\ell}(x) \psi'_{\ell}(mx) }{\mu \psi_{\ell}(mx) \xi'_{\ell}(x)-m\xi_{\ell}(x)\psi'_{\ell}(mx) } 
\end{equation} 
where $ \psi_{\ell}  $, $ \xi_{\ell} $ are Riccati-Bessel functions \cite{DLMF25.12}, and $ \mu $ is the magnetic permeability of the sphere.
The extinction, absorption and  scattering cross-section efficiencies read~\cite{bohren} 
$$ Q_{ext} = Q_{abs}+Q_{scat},  $$
  $$ Q_{abs} = \frac{2}{x^2}  \sum_{\ell=1}^{\infty}{(2{\ell}+1)(Re{[a_{\ell}]}- \vert a_{\ell}\vert ^2 +Re[b_{\ell}]- \vert b_{\ell}\vert^2}), $$
$$ Q_{scat} = \frac{2}{x^2}  \sum_{\ell=1}^{\infty}{(2{\ell}+1)(| a_{\ell}|^2 +|b_{\ell}|^2}). $$
Information on the directionality of the scattered radiation can  be obtained in terms of $ a_{\ell} $ and  $ b_{\ell} $ by defining the differential scattering cross-sections in the forward ($\theta=0$) and
backward ($\theta = \pi)$ directions~\cite{bohren}:
\begin{equation} \frac{dQ}{d\Omega}\bigg\vert_{\theta=0} \hspace{-10pt}= Q_f = \frac{1}{x^2}\left\vert \sum_{\ell=1}^{\infty}{(2\ell+1) (a_{\ell}+ b_{\ell})} \right\vert^2  \label{c1}
\end{equation}
 \begin{equation} \frac{dQ}{d\Omega}\bigg\vert_{\theta=\pi} \hspace{-10pt}= Q_b = \frac{1}{x^2}\left\vert \sum_{\ell=1}^{\infty} {(2{\ell}+1)(-1)^{\ell}(a_{\ell}-b_{\ell})} \right\vert^2. \label{c2}
 \end{equation}
The conditions for ZBS and ZFS were established by Kerker {\it et al.}~\cite{ker} for magnetodielectric spheres as a result of interferences between magnetic and electric scattering resonances~\cite{ker}. When the electric and magnetic multipoles with the same amplitudes oscillate in-phase so that $a_{\ell}  =   b_{\ell} $, ZBS is possible as it can be seen in Eq.~\ref{c2} (first Kerker condition); this occurs provided $\epsilon$ = $\mu$. On the other hand, when the electric and magnetic multipoles oscillate out-of-phase, that is $a_{\ell} = -   b_{\ell}$, ZFS may arise as Eq.~\ref{c1} reveals (second Kerker condition); this occurs when the condition $ \epsilon=\frac{4- \mu}{2\mu+1} $ is satisfied in the quasistatic limit. Note that a complete vanishing of forward scattering cannot occur for any noninvisible passive system since, according to the optical theorem, this would strictly imply in vanishing of absorption and scattering cross sections~\cite{alu2010,lee2017}. However, there is no such restriction for ZBS, which is our interest here, while nearly zero forward scattering is possible for core-shell nanoparticles~\cite{Lee}. In the next subsection we shall demonstrate that ZBS can be achieved in composite, non-magnetic dielectric microspheres.       

\begin{figure}
\includegraphics[width = 3in]{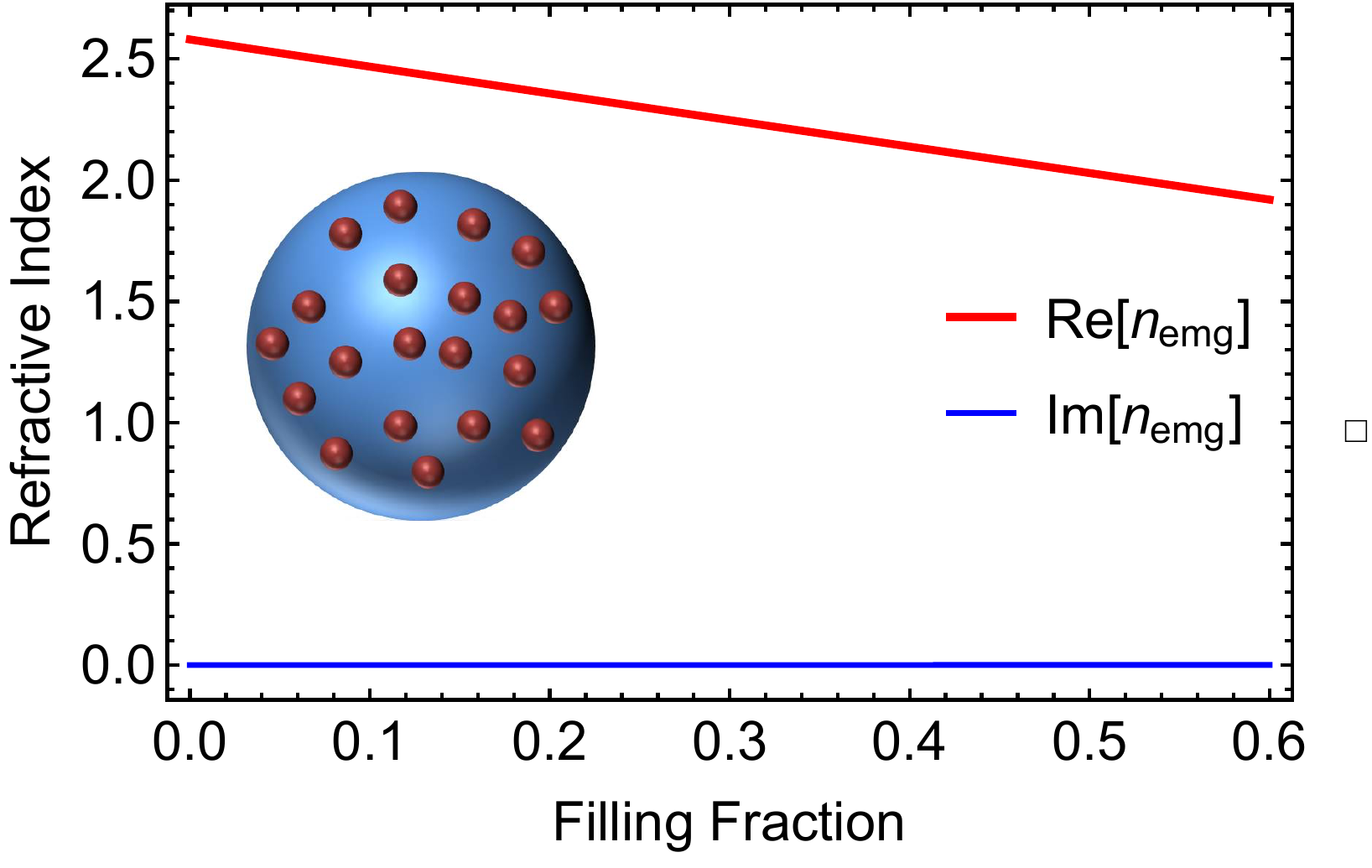} 
\caption{  Real (red) and imaginary (blue) parts of the composite effective refractive index $n_{\rm emg}=\sqrt{\epsilon_{\rm emg}\mu_{\rm emg}},$
calculated with the extended Maxwell Garnett theory, 
versus volume filling fraction.
  The inset
  illustrates the composite made of  ${\rm SiO_2}$ spherical inclusions (radius $8\,{\rm nm}$) distributed inside   a ${\rm SiC}$ sphere.}
\label{fig:1}
\end{figure} 

 \subsection{Extended Maxwell-Garnett Theory }
 
When electromagnetic radiation propagates through a host medium with inclusions (for instance, a porous medium) much smaller than the characteristic wavelength, we can describe such a material by an effective medium theory \cite{Choy_EMT}. Two of the most popular effective medium theories are the so called Maxwell-Garnett (MG) theory \cite{max, max1} and its generalization, the Extended Maxwell-Garnett (EMG) theory \cite{Doyle_89, rup}. In the former one considers
 small spherical particles of radius $a$ and permittivity $\epsilon_{\rm i}$ embedded in host medium of permittivity $\epsilon_{\rm h}$, with a volume filling fraction $f$. Then, MG theory consists essentially in invoking the Lorentz-Lorenz formula \cite{bohren} with the static polarizability of a dielectric sphere, resulting in \cite{rup}  
\begin{equation}
\epsilon_{\rm mg} = \epsilon_{\rm h} \dfrac{\epsilon_{\rm i}(1+2f) + 2\epsilon_{\rm h}(1-f)}{\epsilon_{\rm i}(1-2f) + \epsilon_{\rm h}(2+f)} .
\label{MG}
\end{equation}
The MG effective permittivity (\ref{MG}) is independent of the size of the inclusions, which however must satisfy the condition 
 ${\rm {\it{a}}} \ll \lambda_0.$
 
 \begin{figure}
\includegraphics[width = 3.in]{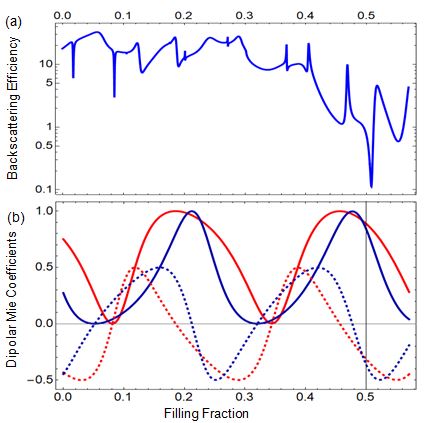}
\caption{(a) Backscattering efficiency $Q_b$ versus volume filling fraction $f,$ showing almost zero backscattering efficiency around $f=0.5$ for a composite microsphere of radius $r=1825\,{\rm  nm}.$  (b) Real and imaginary parts of the dipolar Mie coefficients versus filling fraction. The vertical line at $f=0.5$ shows the intersection of ${\rm Re}(a_1)$ (solid red) and ${\rm Re}(b_1)$ (solid blue), which happens to be quite close to the intersection 
of ${\rm Im}(a_1)$ (dashed red) and ${\rm Im}(b_1)$ (dashed red).}
\label{fig:QBS}
\end{figure}

The extended Maxwell Garnett (EMG) theory  is still based on the Lorentz-Lorenz formula, but now the (electric) dipolar polarizability is given by the full electrodynamical expression, leading to \cite{rup}
\begin{equation}
\epsilon_{\rm emg}=\epsilon_{\rm h}\dfrac{x^3+3if a_1}{x^3-\frac{3}{2}if a_1}
\label{epsEMG}
\end{equation}
where $a_1$ is the electric dipolar  Mie coefficient (\ref{aell}), and $x= \sqrt{\epsilon}_{h} \omega a /c$ is the size parameter  within the host medium. The EMG is most useful when we have $x \ll 1$ (so the inclusions may be considered as dipoles), but the size parameter  within the inclusions
$y = \sqrt{\epsilon}_{i} \omega a /c$  is not necessarily small (so the scattering has to be solved with the Mie theory). By an analogous reasoning, the EMG also gives rise to an effective magnetic permeability \cite{rup}
\begin{equation}
\mu_{\rm emg}=\mu_{h}\dfrac{x^3+3if b_1}{x^3-\frac{3}{2}if b_1}
\label{muEMG}
\end{equation}
where $b_1$ is the magnetic dipolar  Mie coefficient (\ref{bell}). In the framework of EMG theory, even strictly non-magnetic materials may give rise to an effective permeability, as long as $b_1$ is non negligible. We should also stress that these results for $\epsilon_{\rm emg}$ and $\mu_{\rm emg}$ are not necessarily restricted to low filling fractions, provided the positions of the inclusions are uncorrelated \cite{multipolescattering}.
We would like to apply these ideas to the scattering of light from a microsphere of SiC (the host), with nanospheres of ${\rm SiO_2}$ (the inclusions) embedded in it, as shown in the inset of Fig. 1. The effective refractive index $ n_{\rm emg}= \sqrt{\epsilon_{\rm emg} \mu_{\rm emg}}$
as a function of volume filling fraction $f$  is also shown in Fig. 1.
We take $\lambda_0=1064\,{\rm  nm}$, 
the typical wavelength in most optical tweezers setups.
The radius of the microsphere is $r=1825\,{\rm  nm}$ while the nanospheres radii are $a=8\, {\rm nm}$, and the composite is immersed in water. In Fig. \ref{fig:QBS}a the backscattering cross section $Q_b$ is plotted versus volume filling fraction, and we see a big dip right at $f=0.5$. This is partially explained by the plots in Fig. \ref{fig:QBS}b, where it is shown that around $f=0.5$ the $a_1$ and $b_1$ Mie coefficients of the composite microsphere nearly coincide, thus fulfilling the first Kerker condition for ${\ell}=1$ and therefore reducing the backscattering.
 It is possible to show that subsequent electric and magnetic Mie coefficients also meet the first Kerker condition around $f=0.5$, leading to the stark suppression of backscattering shown in Fig. \ref{fig:QBS}(a).


\subsection{Mie-Debye theory of optical tweezers }

Optical trapping of microspheres with a single laser beam is possible only by employing high numerical aperture objectives well beyond the paraxial regime. 
Thus, an accurate theoretical description must be based on
a nonparaxial modelling of the trapping beam. 
The Mie-Debye theory of optical tweezers \cite{epl} combines the
electromagnetic generalization~\cite{RichardsWolf} of Debye's scalar model 
for aplanatic focused beams
with Mie scattering theory. 
Each plane wave component of the incident beam leads to a scattered field component which is easily related to standard Mie-scattering formulae with the help of Wigner 
finite rotation matrix elements $d^{\ell}_{m,m'}(\theta)$ \cite{Edmonds}, where $\theta$ is the angle between each wavevector and the $z$-axis.

Since  the optical force always points towards the beam axis $z$ at any position 
 on the $xy$ plane, it is sufficient to analyze the optical force component $F_z$ along the $z$-axis in order to discuss the requirements for three-dimensional trapping. 
When the sphere center and the trapping beam focal point are aligned along the $z$-axis,  the multipole coefficients for the incident laser beam are given by
\begin{equation}\label{Gl}
 G_{\ell} =  \int_0^{\theta_0} d\theta \sin\theta \sqrt{\cos\theta}  e^{-\gamma^{2} \sin^2 \theta}\, \exp(ik z \cos\theta)\, d_{1,1}^{\ell}(\theta)
 \end{equation}
where $z$ is the position of the sphere center with respect to the focal point, and $\gamma$ is the ratio of the objective focal length to the laser beam waist
at the objective entrance port.
The angular aperture $\theta_0=\sin^{-1}(\mbox{NA}/n)$ is defined by the objective numerical aperture (NA).

It is convenient to define the dimensionless force efficiency \cite{ashkinbio}
\[
Q_z = \frac{F_z}{n P/c}
\]
where $P$ is the laser beam power at the sample region.
We write the force efficiency as the sum of two contributions:
\begin{equation}
Q_z=Q_e{}_z + Q_s{}_z. \label{Q}
\end{equation}
The extinction term  \cite{epl}
\begin{equation}
Q_e{}_z = -\frac{4 \gamma^2}{A} \mbox{Im} \sum^{\infty}_{\ell=1}(2\ell+1) (a_{\ell}+b_{\ell}) G_{\ell} 
\left(\dfrac{\partial G_{\ell}}{\partial (kz)}\right)^*
\end{equation}
accounts for the rate at which linear momentum is removed from the incident laser beam. The factor 
\(
A = 1-\exp(-2\gamma^2\sin^2\theta_0)
\)
is the fraction of the trapping beam power that fills the objective entrance aperture. 

The  term $Q_s{}_z$ in (\ref{Q}) 
 represents the negative of the linear momentum rate carried away by scattered field:
\begin{eqnarray}
&&Q_s{}_z = -\frac{8\gamma^2}{A}   \mbox{Re} \sum^{\infty}_{\ell=1}\left[\frac{\ell(\ell+2)}{(\ell+1)} (a_{\ell} a_{\ell+1}^{*}+b_{\ell} b_{\ell+1}^{*}) G_{\ell} G_{\ell+1}^{*} \right. \nonumber \\
&&\left. \hspace{80pt} +\,\frac{(2\ell+1)}{\ell(\ell+1)} a_{\ell} b_{\ell}^{*}G_{\ell} G_{\ell}^{*} \right] 
\end{eqnarray}

\begin{figure}
\includegraphics[width = 3.in]{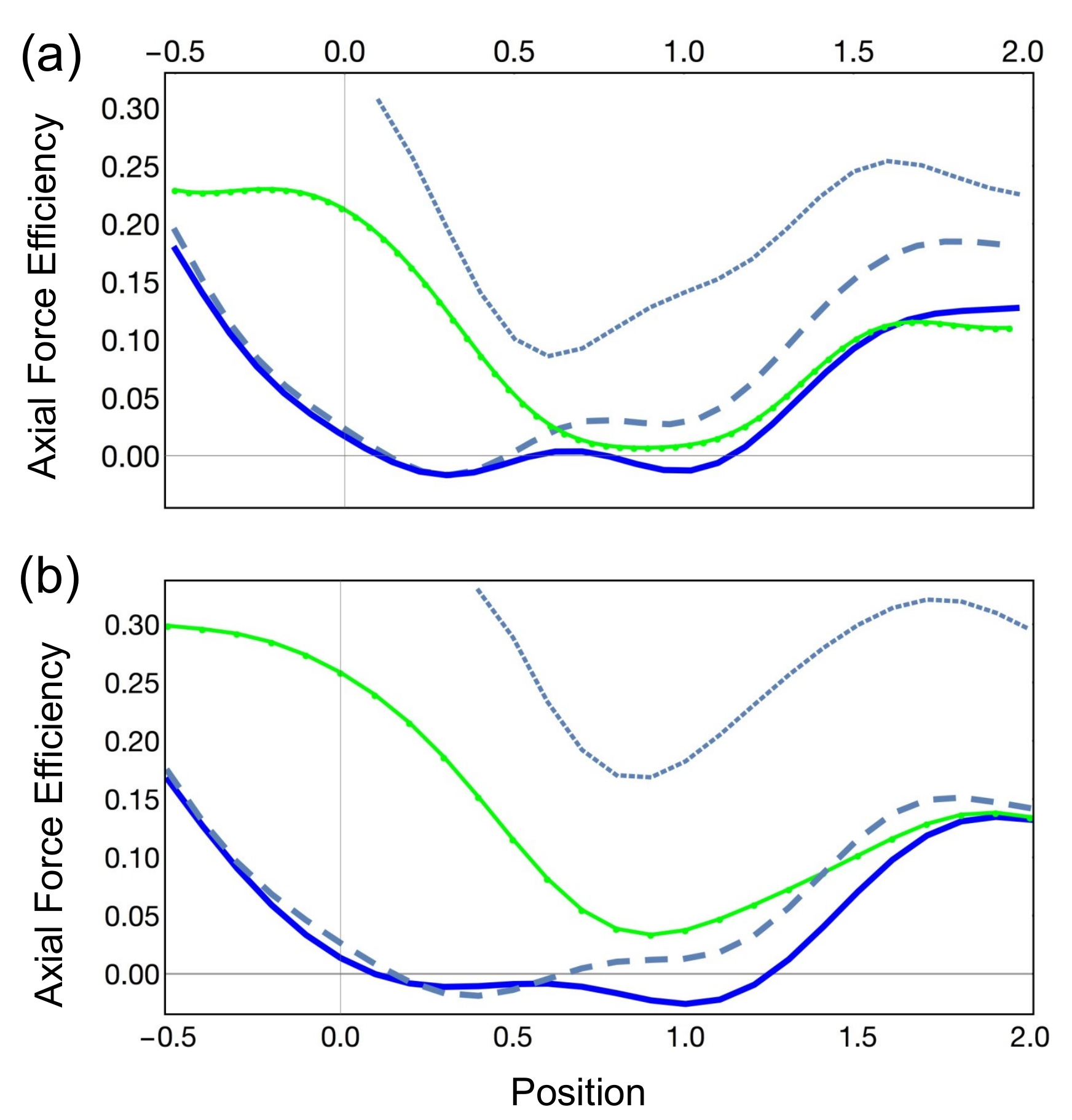}
\caption{ Axial force efficiency $Q_z$ as a function of position (in units of sphere radius) along the laser axis for (a)  sphere radius $r=1825\,{\rm nm}$ with  filling fractions $ f=0$  (dotted line),  $f=0.37$ (dashed line), $f=0.5$ (solid line) and $f=0.55$ (solid-dotted line); and (b)  $r=1500\,{\rm nm}$ with $f=0$  (dotted  line),  $f=0.43$ (dashed line), $f=0.5$ (solid-dotted line), and  $f=0.58$ (solid line). }
 \label{fig: 3}
\end{figure}

 Most optical tweezers setups employ oil-immersion objectives. In this case,  the 
 refractive index mismatch at the glass-water interface leads to spherical aberration of the focused trapping beam, which can be taken into account 
 by introducing suitable phase factors \cite{Torok} in the expression
 (\ref{Gl}) for  the multipole coefficients $G_{\ell}.$
 The resulting Mie-Debye-spherical aberration (MDSA) theory of optical tweezers 
  \cite{Viana07} thus
  accounts for the spherical aberration introduced by refraction at the glass slide. 
  Excellent blind agreement with experiments was obtained 
  when additional optical aberrations are also included  \cite{Dutra12,Dutra14}.


\section{Results and Discussion}

Figure \ref{fig: 3}a shows the optical axial force $Q_z$ 
variation with the microsphere position $z$ with respect to the focal point (normalized by the radius $r=1825\, {\rm nm}$).
The parameters used in numerics are listed in Ref.~\cite{parameter}. 
Figure \ref{fig: 3}a shows  that an homogeneous ${\rm SiC}$ sphere ($f=0$) of this size cannot be optically trapped since 
$Q_z>0$ for all values of $z.$ 
In fact, radiation pressure provides the dominant contribution to the optical force 
due to the large refractive index of ${\rm SiC},$
thus leading to an optical force pointing along the laser propagation direction for all values for the sphere position. 
 On the other hand,
 the presence of ${\rm SiO_2}$ inclusions allows for optical trapping
 since there are stable equilibrium positions  ($Q_z=0$) when 
 considering $f=0.37$ and $f=0.5,$
corresponding to the dashed and solid lines in  Fig.~\ref{fig: 3}a.

\begin{figure} 
\includegraphics[width = 3.4in]{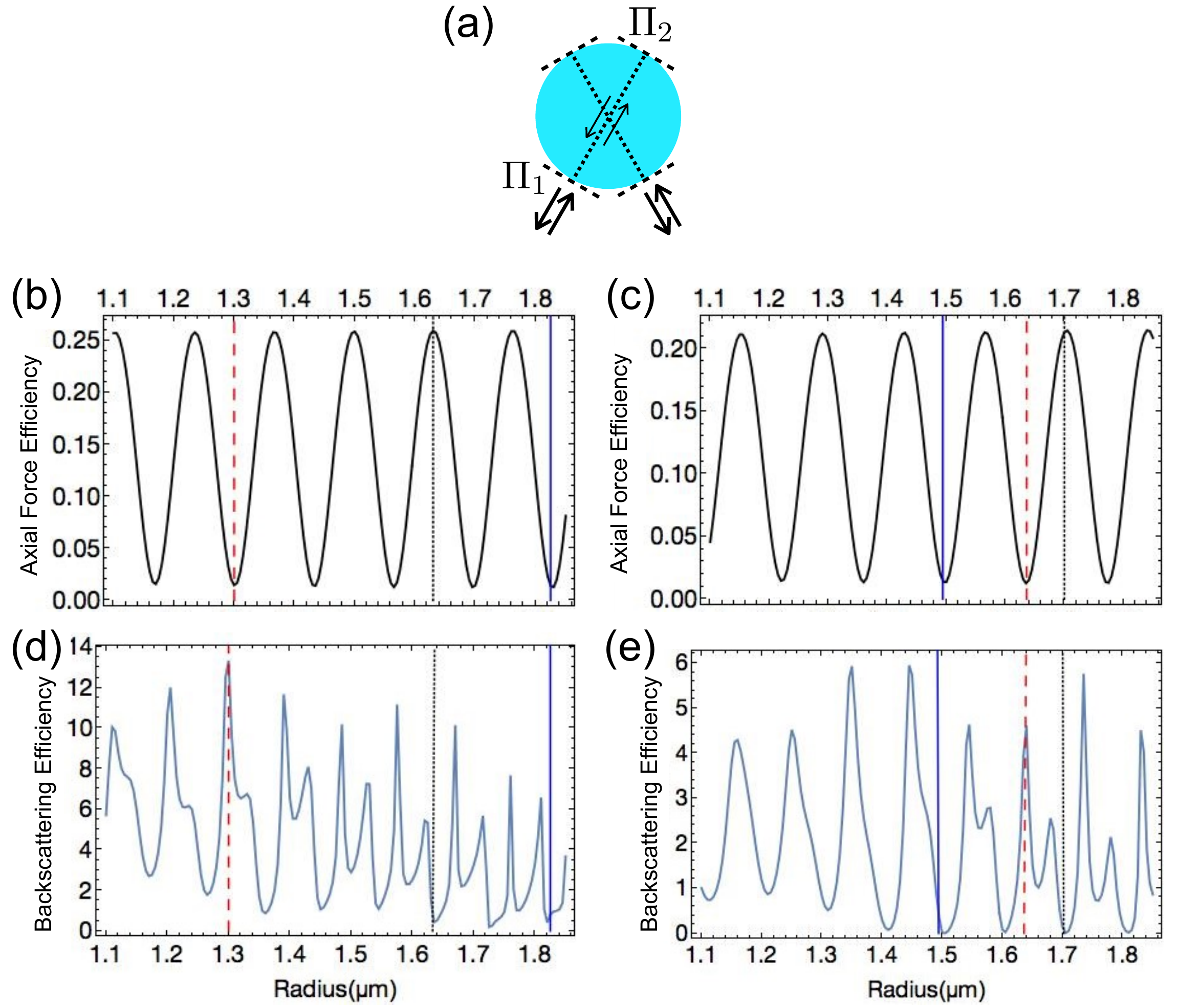}
\caption{
(a) Scheme representing Mie scattering when the sphere center is located at the focal point of the incident beam. In this case, the resulting optical force
\mbox{$Q_z(z=0)$}, plotted as a function of the sphere radius $r$ in (b) for $f=0.5$ and in (c) for $f=0.58,$
 can be interpreted in terms of the interference between the fields reflected at planes $\Pi_1$ and $\Pi_2.$ 
(d,e) 
 Backscattering efficiency $Q_b$ versus $r$ for   (d)
 $f=0.5$ and (e) $f=0.58$. 
 The vertical lines indicate the filling fractions used in Figs. 5 and 6. Optimal trapping is achieved by selecting $r$ to be a simultaneous minimum of both $Q_z$ and $Q_b$ (solid vertical lines). }
\label{fig:4}
\end{figure}

The latter case corresponds to the situation where the first Kerker condition $Q_b=0$ is fulfilled.
 Interestingly, this case corresponds precisely to the situation where trapping is more stable, {\it i.e.} where the range of position values for which 
 the optical force is in opposition to the propagation direction
 ($Q_z<0$) is larger, as indicated by the solid line in Figure \ref{fig: 3}a. Besides, here it is important to emphasize that the effect of 
 making the trap more stable
 is not a trivial consequence of  decreasing the effective refractive index by increasing the relative amount of ${\rm SiO_2}$ in the composite. 
 Indeed, this can be seen by comparing the curves of $Q_z$ for two slightly different values of the filling fraction, $f=0.5$ and $f=0.55$. While the former leads to optimal optical trapping, in the latter case optical trapping is impossible despite the fact that it corresponds to a larger density of inclusions, and hence to a smaller effective refractive index as shown in Fig. \ref{fig:1}.
  This results unveils the crucial role played by minimizing backscattering, achieved at the first Kerker condition as discussed  in connection with Fig. \ref{fig:QBS}.
  We identify such configuration as providing  the optimal and more stable optical tweezing.  All previous results are robust against varying the host sphere radius, as illustrated in Fig. \ref{fig: 3}b, where an analogous calculation is carried out considering $r=1500\, {\rm nm}.$
   The first Kerker condition is now satisfied at $f=0.58$, and the  findings are qualitatively the same, except that now there is a single equilibrium position with an enlarged stability range along the $z$ axis, instead of two stable equilibrium positions as  in the case considered in  Fig. \ref{fig: 3}a.

\begin{figure} 
\includegraphics[width = 3.2in]{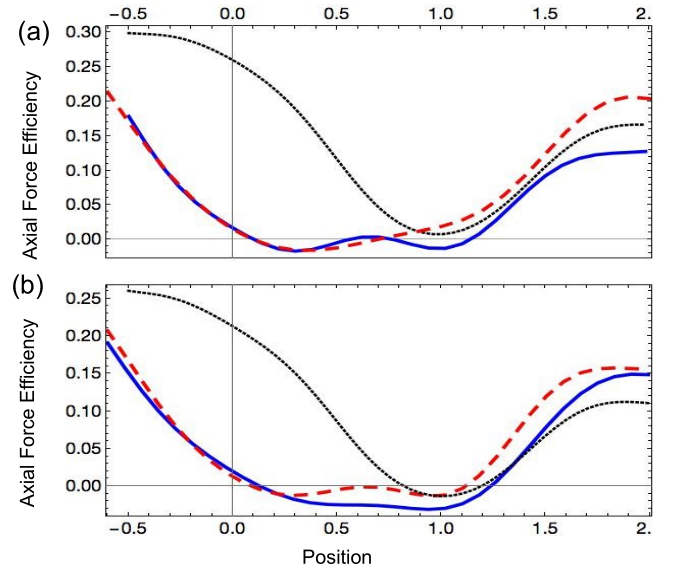}
\caption{Axial force efficiency $Q_z$ as a function of position (in units of sphere radius) along the laser axis for (a) $f=0.5$ with radii
$r=1300\,{\rm nm}$ (dashed line), 
 $1630\,{\rm nm}$ (dotted line), and $1825\,{\rm nm}$ (solid line); and for (b) $f=0.58$ 
 with radii $r=1490\,{\rm nm}$ (solid line), 
 $1636\,{\rm nm}$ (dashed line), and $1700\,{\rm nm}$ (dotted line). Such radii correspond to minima of $Q_z(z=0),$ $Q_b$ or both.  
}  
\label{fig:5}
 \end{figure}

\begin{figure} 
\includegraphics[width = 3.2in]{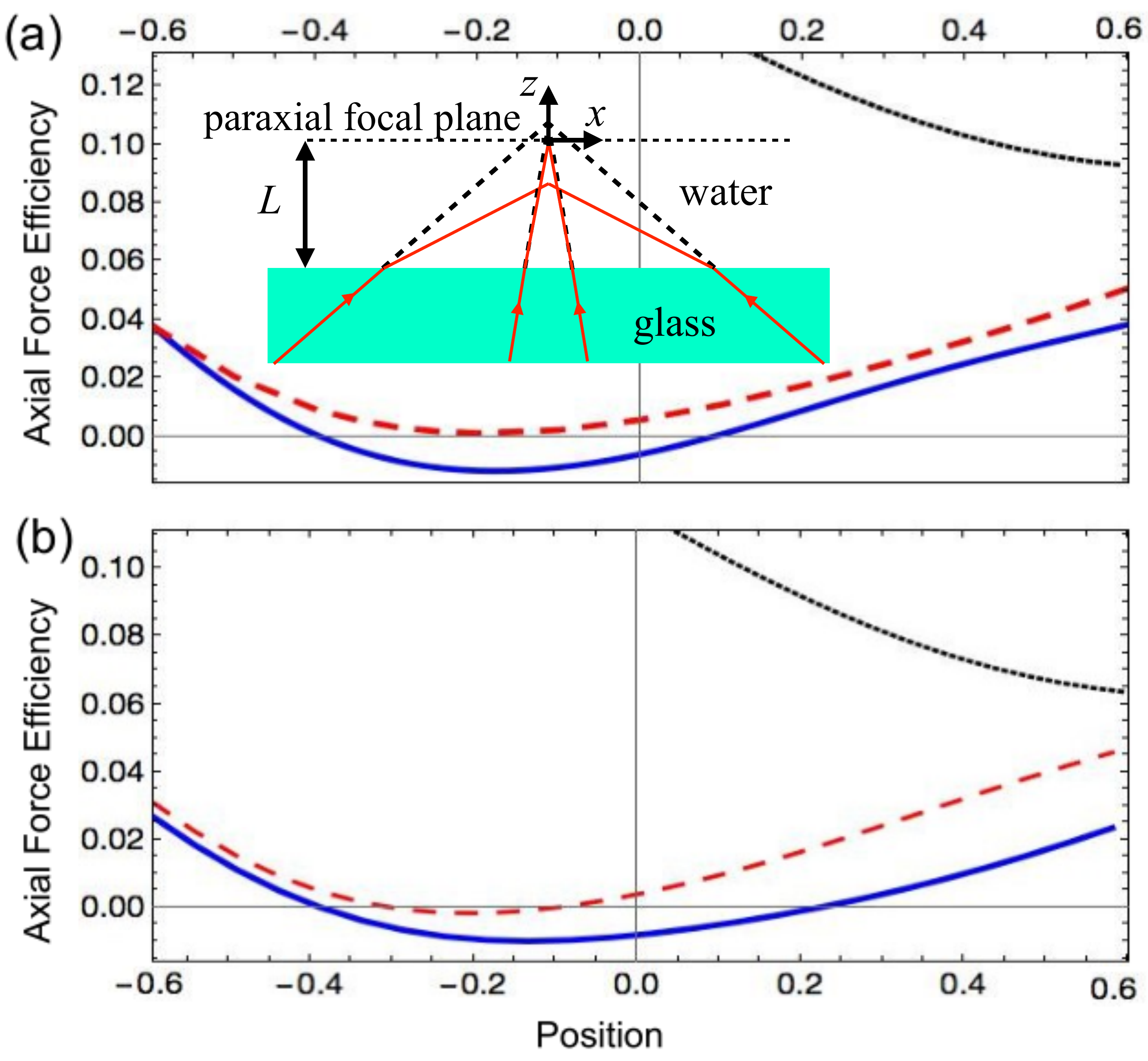}
\caption{Same conventions as in Fig.~5,  with the spherical aberration introduced by the 
glass-water interface taken into account. Here $z=0$ represents the paraxial focal point, 
which is at a distance $L=7 r$ above the glass slide, as illustrated by the inset. 
The point of maximum energy density (diffraction focus) is located in between the 
paraxial focus and the glass slide ($z<0$).}  
\label{fig:6}
\end{figure}

In order to 
gain further insight into the conditions needed for optimal trapping, 
we analyze 
 the optical force  $Q_z$ 
 when the sphere center is located at the focal point
$(z=0).$
 In this case, 
 all optical rays forming the incident laser beam propagate along radial directions and hence are not deflected when refracting into the microsphere. In other words, 
  all rays composing the incident  beam correspond to a vanishing 
 impact parameter since they intersect at the sphere center,  as illustrated by Fig.~\ref{fig:4}a.
We calculate $Q_z(z=0)$ as a function of the sphere radius $r$
for  $f=0.5$ (Fig.~\ref{fig:4}b) and $f=0.58$ (Fig.~\ref{fig:4}c). 
Our results obtained within the wave-optical Mie-Debye theory can be interpreted as resulting from 
 the  interference between the field reflected at the water-composite interface
  (plane $\Pi_1$ in Fig.~\ref{fig:4}a)
   and the one reflected at the composite-water interface  (plane $\Pi_2$) after 
   a round-trip propagation along the sphere diameter \cite{epl}.
   Indeed, the minima shown in  Figs.~\ref{fig:4}(b,c) are well approximated by 
   \begin{equation} \label{minima}
   r_{\rm min}= j\, \frac{\lambda_0}{4n_{\rm emg}}, \,j \in \, \mbox{integers}
      \end{equation}
  with
   $\lambda_0/(4n_{\rm emg})\approx 131\,{\rm nm}$ ($f=0.50$) and  $136\,{\rm nm}$ ($f=0.58$), as in 
   a slab of thickness $2r$ and refractive index $n_{\rm emg}.$
     When (\ref{minima}) is met, the two fields interfere destructively, 
  the reflectivity  
  is minimized and so is the resulting radiation pressure, thus favoring trapping. 
  
 The backscattering efficiency $Q_b$ for a plane wave cannot be interpreted along the same lines. Within ray optics,
 an incident plane wave 
 corresponds to parallel optical rays 
which are deflected at the water-composite interface along various angles depending on the impact parameter. Hence 
the variation of $Q_b$ versus $r$ shown in 
 Figs.~\ref{fig:4}d ($f=0.50$) and \ref{fig:4}e ($f=0.58$) 
 brings a complementary information on trapping optimization. The first Kerker condition corresponds to the radii 
 giving a minimum $Q_b.$
 The optimal trapping conditions discussed in connection with Fig.~\ref{fig: 3}a correspond to picking
 $r=1825\,{\rm nm},$ which is 
  a simultaneous minimum of both 
 $Q_z(z=0)$ and $Q_b$ 
as indicated by
  vertical solid lines in Figs.~\ref{fig:4}b and \ref{fig:4}d.
  The additional vertical lines 
  at $r=1300\,{\rm nm}$ and $r=1630\,{\rm nm}$
  indicate a minimum  and a maximum of  $Q_z(z=0)$ which turn out to be a maximum and a minimum of  $Q_b,$ respectively.  
   In Fig.~\ref{fig:5}a, we compare those two configurations with the optimal one by plotting the corresponding
   variations of the axial force versus position (in units of sphere radius).  For the former case (dashed line), radiation pressure is minimized near the focal point, 
 but then builds up as the sphere is displaced along the optical axis. As a consequence, the resulting trapping range is significantly smaller. 
 On the other hand, for $r=1630\,{\rm nm}$ (dotted line), no trapping is possible since radiation pressure is maximized near the focal point due to constructive interference of the reflected field components shown in Fig.~\ref{fig:4}a. A similar comparison is illustrated by Fig.~\ref{fig:5}b for the case $f=0.58,$ but now the maximum of $Q_z(z=0)$ satisfying the first Kerker condition
($r=1700\,{\rm nm}$, dotted line)  does lead to trapping, although at a position far above the focal point, and over a shorter stability range. Taken together, Figs.~\ref{fig: 3} and \ref{fig:5} show that selecting a microsphere satisfying the first Kerker condition clearly helps to increase the trap stability. Nevertheless, the discussion above indicates that the requirement
(\ref{minima}) for  destructive interference at the focal point is more important for trapping high-refractive index 
microspheres with the ideal, aplanatic (aberration-free) trapping beams considered so far.  
 
 The first Kerker condition  becomes comparatively
  more important when spherical aberration is taken into account. Indeed, 
the properties of optical tweezers are extremely sensitive 
to small quantities of optical aberrations \cite{Dutra12, Dutra14}. A very common source of spherical aberration is the refractive index mismatch between 
the glass slide and water-filled sample when employing high-NA oil-immersion objectives \cite{Torok}. As discussed in the end of Sec. II.C, 
the effect of the glass-water interface on the trapping beam is taken into account within the Mie-Debye-spherical aberration (MDSA) theory of optical tweezers \cite{Viana07}. 
In Fig.~\ref{fig:6}, we plot  the MDSA results for the axial force efficiency
 variation with the microsphere position, with the same parameters and conventions employed
in Fig.~\ref{fig:5}. We take the paraxial focus at a distance $L=7\,r$, well above the glass slide, so that the optical reverberation between the glass slide and the 
sphere is negligible~\cite{Dutra2016}. As illustrated in the inset of Fig.~\ref{fig:6}a, 
the energy density is distributed between the paraxial focus ($z=0$) and the glass slide,
 and the diffraction focus is located below the paraxial one. As a consequence, 
the stable equilibrium positions shown in Fig.~\ref{fig:6} are now located at $z<0.$ 
The comparison between Figs~\ref{fig:5} and \ref{fig:6} shows that the trap stability range is 
shorter and the maximum restoring counter-propagating force is weaker, as expected since optical aberrations degrade the focal region leading to a reduction of the 
energy density gradients.

More importantly, Fig.~\ref{fig:6} shows
that our trapping proposal, 
with the radii and filling fractions
corresponding to the solid lines in  Figs.~\ref{fig:5} and \ref{fig:6},
  still works even in the presence of a significant amount of spherical aberration. 
The interpretation in terms of a parallel-planes interferometer (see Fig.~\ref{fig:4}a) now holds only
for the paraxial Fourier components of the incident beam, which provide a significant fraction of the total radiation pressure. Thus, selecting a radius 
satisfying the condition
(\ref{minima})
for the ideal aplanatic case is still helpful but  no longer sufficient 
to achieve trapping, as illustrated by the dashed lines in Fig.~\ref{fig:6}. In fact, trapping is now possible only by 
 picking a 
 radius from (\ref{minima})
 that also turns out to minimize $Q_b,$ 
 as indicated by the solid vertical lines in Fig.~\ref{fig:4}.

\section{Conclusion}

Recent advances in the field of metamaterials have made possible 
the achievement of highly directional Mie scattering in the optical domain. 
We have shown that the condition of zero backward scattering (first Kerker condition) can be applied to 
optically trap high-index microspheres with a single focused laser beam, thereby widening the range of applications 
of this popular experimental technique. 

We have considered SiC microspheres with ${\rm SiO_2}$ inclusions immersed in water.
 We have shown that backward scattering can be strongly suppressed by tuning 
the filling fraction. In such cases, stable trapping is possible provided that the sphere radius and filling fraction are such as to also lead to a  destructive interference between the fields reflected at the external 
and internal interfaces illustrated by Fig.~\ref{fig:4}a. 

Fulfilling the first Kerker condition considerably enhances stability, but 
is not a necessary condition for 
trapping when optical aberrations are disregarded. On the other hand, for  typical setups with oil-immersion high-NA objectives, 
it is essential to satisfy the Kerker condition in order to achieve trapping of high-index microspheres, 
for the refractive index mismatch between the glass slide and the water-filled sample introduces a 
significant amount of spherical aberration. 

\begin{acknowledgments}
We thank D. S. Ether jr and N. B. Viana for inspiring
discussions.  This work has been supported by 
 the Brazilian agencies National Council for Scientific and Technological Development (CNPq), the
National Institute of Science and Technology Complex Fluids (INCT-FCx), the
Carlos Chagas Filho Foundation for Research Support of Rio de Janeiro
(FAPERJ) and the S\~ao Paulo Research Foundation (FAPESP).
F. A. P. thanks the Royal Society -- Newton Advanced Fellowship (grant no. NA150208) for financial support.

\end{acknowledgments}


\begin{thebibliography}{99} 

\bibitem{ashkin1986} A. Ashkin, J. M. Dziedzic, J. E. Bjorkholm, and Steven Chu, Observation of a single-beam gradient force optical trap for dielectric particles. Opt. Lett. { \bf 11}, 288 (1986).
\bibitem{ashkin2006} A. Ashkin, {\it Optical Trapping and Manipulation of Neutral Particles using Lasers: A Reprint Volume With Commentaries}
(World Scientific, Singapore, 2006).
\bibitem{fazal2011} F. M. Fazal and S. M. Block, Optical tweezers study life under tension.  Nat.  Photonics {\bf 5}, 318 (2011). 
\bibitem{toyabe2010} S. Toyabe, T. Sagawa, M. Ueda, E. Muneyuki and M, Sano,  Experimental demonstration of information-to-energy conversion and validation of the generalized Jarzynski equality.  Nat.  Physics {\bf 6}, 988 (2010).
\bibitem{brut2012}  A. Berut, A.  Arakelyan, A. Petrosyan,  S. Ciliberto,    R. Dillenschneider  and    E. Lutz, Experimental verification of Landauer's principle linking information and thermodynamics.  Nature {\bf 483}, 187 (2012).
\bibitem{diney}D. S. Ether jr., L. B. Pires, S. Umrath, D. Martinez, Y. Ayala, B. Pontes, G. R. de S. Araujo, S. Frases, G. L. Ingold, F. S. S. Rosa, N. B. Viana, H. M. Nussenzveig and P. A. Maia Neto,
 Probing the Casimir force with optical tweezers. Europhys.  Lett. {\bf 112},  44001 (2015). 
 

\bibitem{ker} M. Kerker, D.-S.  Wang,  C. L. Giles,  Electromagnetic scattering by magnetic spheres.   J. Opt. Soc.  Am.  \textbf{73}, 765 (1983).

\bibitem{monticone} F. Monticone and A. Al\'u, The quest for optical magnetism: from split-ring resonators to plasmonic nanoparticles and nanoclusters.  J. Mat. Chem. C {\bf 2}, 9059 (2014).



\bibitem{Si}  S. Person, M. Jain,  Z. Lapin, J. J.  S\'aenz, G. Wicks, and L. Novotny,  Demonstration of zero optical backscattering from single nanoparticles.  Nano Lett.  \textbf{13}, 1806 (2013).

\bibitem{fu2013} Y. H. Fu, A. I. Kuznetsov, A. E. Miroshnichenko, Y. F. Yu, and B. Luk'yanchuk,   Directional visible light scattering by silicon nanoparticles.  Nat. Commun. {\bf 4}, 1527 (2013).

\bibitem{ge}  A. Garc\'{\i}a-Etxarri, R. G. Medina, L. S. F. P\'erez, C. L\'opez, L. Chantada, F. Scheffold, J. Aizpurua, M. N. Vesperinas, and J. J. S\'aenz,  Strong magnetic response of submicron silicon particles in the infrared.  Opt. Express  \textbf{19}, 4815 (2011).

\bibitem{geffrin2012} J. M. Geffrin, B. G. C‡\'amara, R. G. Medina, P. Albella,
L. S. F. P\'erez, C. Eyraud, A. Litman, R. Vaillon and  F. Gonz\'alez, Magnetic and electric coherence in forward- and back-scattered electromagnetic waves by a single dielectric subwavelength sphere. Nat.  Commun. {\bf  3}, 1171 (2012).

\bibitem{alaee2015} R. Alaee, R. Filter, D. Lehr, F. Lederer, and C. Rockstuhl, A generalized Kerker condition for highly directive nanoantennas. Opt. Lett. {\bf 40}, 2645 (2015).


\bibitem{li2015} Y. Li, M. Wan, W. Wu, Z. Chen, P. Zhan, and Z. Wang,  Broadband zero-backward and near-zero-forward scattering by metallo-dielectric core-shell nanoparticles. Sci. Rep. {\bf 5}, 12491 (2015).

\bibitem{xie2015} Y.-M. Xie, W. Tan, and Z.-G. Wang,  Anomalous forward scattering of dielectric gain nanoparticles. Opt. Express  {\bf 23}, 2091 (2015).


\bibitem{liu2012} W. Liu, A. E. Miroshnichenko, D. N. Neshev, and Y. S. Kivshar, Broadband unidirectional scattering by magneto-electric core shell nanoparticles. ACS Nano {\bf 6}, 5489 (2012).


\bibitem{kuznetsov2016} A. I. Kuznetsov, A. E. Miroshnichenko, M. Brongersma, Y. S.
Kivshar, and B. Luk'yanchuk, Optically resonant dielectric nanostructures.  Science {\bf 354}, 2472 (2016).

\bibitem{rodriguez2014} S. R. K. Rodriguez, F. B. Arango, T. P. Steinbusch, M. A. Verschuuren, A. F. Koenderink, and J. G. Rivas, Breaking the symmetry of forward-backward light emission with localized and collective magnetoelectric resonances in arrays of pyramid-shaped aluminum nanoparticles. Phys. Rev. Lett. {\bf 113},  247401 (2014).



\bibitem{bohren} C. F. Bohren and D. R. Huffman, {\it Absorption and Scattering  of Light by Small Particles}  (Wiley, New York, 1998).


\bibitem{DLMF25.12} NIST digital library of mathematical functions. \url{http://dlmf.nist.gov/25.12}, Release 1.0.16 of 2017-09-18.
F. W. J. Olver, A. B. Olde Daalhuis, D. W. Lozier, B. I. Schneider, R. F. Boisvert, C. W. Clark, B. R. Miller,
and B. V. Saunders, eds.

\bibitem{alu2010} A. Al\'u and N. Engheta, How does zero forward-scattering in magnetodielectric nanoparticles comply with the optical theorem?. J. Nanophoton. {\bf 4}, 041590 (2010).
\bibitem{lee2017} J.Y. Lee, A. E. Miroshnichenko, and R.-K. Lee, Reexamination of Kerker's conditions by means of the phase diagram.  Phys. Rev. A {\bf 96}, 043846 (2017).

\bibitem{Lee} Y.  Tsuchimoto, T. Yano, T. Hayashiand M. Hara, Fano resonant all-dielectric core-shell nanoparticles with ultrahigh scattering directionality in the visible region. 
Opt. Express, \textbf{24}, 14452  (2016). 


\bibitem{Choy_EMT} T. C. Choy, {\it Effective Medium Theory: Principles and Applications}  (Oxford University Press, 2015).
\bibitem{max}  J.  C.  M.  Garnett,  Colours in metal glasses and in metallic films.  Philos. Trans. R. Soc. A  \textbf{203}, 385  (1904).
\bibitem{max1}   J.  C.  M.  Garnett,  Colours in metal glasses, in metallic films and in metallic solutions-II.  Philos. Trans. R. Soc. A \textbf{205}, 237 (1906).
\bibitem{Doyle_89} W. T. Doyle,  Optical properties of a suspension of metal spheres. Phys. Rev. B {\bf 39}, 9852 (1989).
\bibitem{rup} R. Ruppin,  Evaluation of extended Maxwell-Garnett theories. Opt. Commun. \textbf{182},  273  (2000).

\bibitem{multipolescattering} P.  Mallet, C.  A.  Guerin and A.  Sentenac,  Maxwell-Garnett mixing rule in the presence of multiple scattering: derivation and accuracy. Phy. Rev. B \textbf{72}, 014205 (2005).

\bibitem{epl}  P.  A.  Maia Neto and H. M. Nussenzveig, Theory of optical tweezers.  Europhys. Lett. \textbf{50},  702  (2000).
\bibitem{RichardsWolf} B. Richards and  E. Wolf, Electromagnetic diffraction in optical systems II. Structure of the image field in an aplanatic system.
 {Proc. R. Soc. London} A  \textbf{253}, 358 (1959).
\bibitem{Edmonds}  A. R. Edmonds,  {\it Angular Momentum in Quantum Mechanics} (Princeton University Press, 1957).

\bibitem{Torok} P. T\"or\"ok, P. Varga, Z. Laczik, and G. R. Booker, Electromagnetic diffraction of light focused through a planar interface between materials of mismatched refractive indices: an integral representation.
 J. Opt.  Soc.  Am.  A  \textbf{12}, 325  (1995).

\bibitem{Viana07} N. B. Viana, M. S. Rocha, O. N. Mesquita, A. Mazolli, P. A. Maia Neto and H. M. Nussenzveig,
 Towards absolute calibration of optical tweezers.  Phys. Rev. E \textbf{75}, 021914  (2007).
 \bibitem{Dutra12}  R. S. Dutra, N. B. Viana, P. A. Maia Neto and H. M. Nussenzveig,  Absolute calibration of optical tweezers including aberrations.
 Appl. Phys. Lett. \textbf{100}, 131115 (2012).
\bibitem{Dutra14} R. S. Dutra, N. B. Viana, P. A. Maia Neto and H. M. Nussenzveig,  Absolute calibration of forces in optical tweezers. 
Phys. Rev. A  \textbf{90}, 013825 (2014).

\bibitem{parameter}  We have taken the following values for the refractive indexes: 
$n_w=1.332$ (water),  $n_{\rm SiC}=2.59,$  $n_{\rm SiO_2}=1.45,$   $n= 1.51$ (glass). In addition, we have used the following values for the parameters 
defined in Sec.~II.C:
 $\omega_0=2.84$ (beam waist), $\gamma=1.226$ and NA$=1.3.$  

\bibitem{ashkinbio}  A. Ashkin,  Forces of a single-beam gradient laser trap on a dielectric sphere in the ray optics regime.  Biophys. J.  \textbf{61},  569 (1992).

\bibitem{Dutra2016} R. S. Dutra, P. A. Maia Neto, H. M. Nussenzveig, and H. Flyvbjerg, 
Theory of optical-tweezers forces near a plane interface.
Phys. Rev. A  \textbf{94} ,  053848 (2016).




\end{thebibliography}
\end{document}